\documentclass[twocolumn,prl,superscriptaddress,showpacs,preprintnumbers,amsmath,amssymb]{revtex4-1}
\usepackage{graphicx}
\usepackage{color}

\begin{document}

\title{Discontinuous shear modulus determines the glass transition temperature}

\author{Christian L. Klix}
\affiliation{University of Konstanz, D-78457 Konstanz, Germany}
\author{Georg Maret}
\affiliation{University of Konstanz, D-78457 Konstanz, Germany}
\author{Peter Keim}
%\correspondingauthor{Peter.Keim@uni-konstanz.de}
\email{peter.keim@uni-konstanz.de}
\affiliation{University of Konstanz, D-78457 Konstanz, Germany}

\date{\today}

\begin{abstract}
{A solid - amorphous or crystalline - is defined by a finite shear modulus while a fluid lacks such. We thus experimentally investigate the elastic properties of a colloidal glass former near the glass transition: spectroscopy of vibrational excitations yields the dispersion relations of longitudinal and transverse phonons in the glassy state. From the long wavelength limit of the dispersion relation we extract the bulk and the shear modulus. As expected, the latter disappear in a fluid and we measure a clearly resolved discontinuous behaviour of the elastic moduli at the glass transition. This not only determines the transition temperature $T_G$ of the system but also directly addresses recent discussions about elasticity during vitrification. We show that low frequency excitations in our system are plane waves such that continuum elasticity theory can be used to describe the macroscopic behaviour.}
\end{abstract}

%\pacs{05.70.Fh, 05.70.Ln, 64.60.Q-, 64.70.pv, 82.70.Dd}

\maketitle

In general, there are several ways to define glassy systems \cite{Angell1988,Debenedetti2001}. For instance, a characteristic change in the thermodynamic properties volume and enthalpy upon cooling a fluid may be used to define the glass transition temperature $T_G$. Another property often investigated for molecular or atomic glasses is the viscosity $\eta$. Taking into account the rapid slowing down of dynamics, the glassy state is reached when the viscosity exceeds $10^{12}$ Pa$\cdot$s upon crossing the freezing temperature. However, all of these definitions are difficult to apply to soft matter systems. In studies on the mesoscopic or microscopic scale, access to the macroscopic property $\eta$ often is difficult, if not impossible. Secondly, these systems are ``softer'' up to a factor of $10^{-12}$ (2D) or $10^{-15}$ (3D) which rules out the idea of an universal, viscosity-dependent categorization. Thus, other ways have been suggested to pin down $T_G$ \cite{Debenedetti2001}. A manifestation of solidity is the emergence of rigidity, expressed by the shear modulus $\mu$ becoming finite. Microscopically, prerequisites for solidity are long ranged dynamical correlations. While it is easy to understand how such long range correlations emerge in a crystal (based on the concept of long ranged periodic order), solidity is much less easy to grasp on an amorphous, disordered background. As the shear modulus is zero in fluids, we expect a significant change of $\mu$ at the onset of vitrification \cite{Jeong1987,Barrat1988}. Although $\mu$ is a macroscopic property, it is measurable locally via the equipartition theorem which makes this concept well suited for soft matter systems \cite{Gruenberg2004,Reinke2007}. From an experimental point of view, this approach is advantageous because it relies on measuring the change of a quantity from zero to a finite value. On the other hand, the usual approach of measuring the divergence of dynamical quantities like the viscosity is much harder to realize experimentally because it requires access to a much larger parameter space. \\

Elasticity of complex fluids and their viscoelastic behaviour has been tackled mostly by rheological measurements so far \cite{Mason1995}. In general, these methods allow the study of frequency dependent storage $G'(\omega)$ and loss $G''(\omega)$ modulus. The former describes the elastic contribution, while the latter covers dissipative processes upon shear. Thus, a non-vanishing $G'(\omega)>G''(\omega)$ in the limit $\omega\rightarrow 0$ indicates a solid. These quantities have been investigated extensively by experiment as well as theory \cite{Crassous2008,Crassous2013}, and from the data, a clear distinction between solid and fluid state might be drawn. This is also true for flow curves, which might be not only acquired by conventional rheometers but also by microfluidic rheology \cite{Nordstrom2010}. Here we address solely the elasticity by internal fluctuations without external driving. \\

While it has been shown recently that the zero-frequency shear modulus $\lim_{\omega\rightarrow 0} G'(\omega) = \mu$ is a good quantity to observe the glass transition \cite{Klix2012}, there has been an ongoing debate on the behaviour of $\mu$ when an amorphous solid forms. Replica theory approaches \cite{Yoshino2010} predict a continuous growth from zero upon cooling. This agrees with the view from granular systems, where theory predicts critical fluctuations close to the jamming transition of these athermal systems which also lead to an algebraic growth of $\mu$ from zero \cite{OHern2003}. On the other hand, another recent replica theory calculation reveals a discontinuous jump at the glass transition \cite{Szamel2011,Yoshino2012,Yoshino2014}. This complies with mode coupling theory predictions \cite{Leutheusser1984} where $\mu$ is connected to a non-ergodicity parameter appearing in a stress auto-correlation function. A discussion about the discontinuous elasticity in the context of Random First-Order Transition is given in \cite{Biroli2012}. In addition, simulations of hard discs also showed a discontinuous onset of the shear modulus \cite{Klix2012} by accessing the system with the equipartition theorem. Similar results from novel four-point correlation calculations back this up further \cite{Flenner2015}. \\

Here, we evaluate microscopy data from a two dimensional colloidal model system with soft interactions in equilibrium. This gives us access to information on the single particle scale as opposed to atomic or molecular systems, where only ensemble data may be studied. By determining the displacement field and subsequent analysis in Fourier space, the dispersion relations of acoustic-like excitations are obtained and provide insight into the macroscopic elastic properties of the system. As the system is cooled, we are able to confirm the discontinuous nature of the shear modulus at the glass transition.

\section{Methods}

Video microscopy provides access to the full phase space information. We record the trajectories of about 2300 colloidal particles confined to two dimensions at a flat water-air interface. The species A (diameter $\sigma = 4.5\,\mu m$) and B ($\sigma = 2.8\,\mu m$) have a relative concentration of $\xi = N_B/(N_A+N_B)\approx 50\,\%$ where $N_A$ and $N_B$ are the number of particles of both species in the field of view. This prevents the system from crystallization. \\
The superparamagnetic nature of the particles lets us control the particle interactions \textit{in situ} by an external magnetic field $H$. This is expressed by the dimensionless system parameter
\begin{equation}\label{gamma}
	\Gamma = \frac{\mu_0}{4\pi} \cdot \frac{H^2 \cdot (\pi n)^{3/2}}{k_BT}(\xi \cdot \chi_B + (1-\xi)\cdot \chi_A)^2 \, ,
\end{equation}
which effectively acts as an inverse temperature. Here, $n$ denotes the area density and is computed via a Voronoi tessellation. $\chi_{A,B}$ represent the susceptibilities of species A and B, respectively. After equilibration at low $\Gamma$, the system was cooled down stepwise. For different states up to $3.8\cdot 10^5$ snapshots were analyzed. With a frame rate of $\approx 2\,\mathrm s^{-1}$, sampling times up to some $10^5$ seconds were achieved. This is sufficiently long to probe dynamics even in the relaxation time regime $\tau_\alpha$ for highest system parameters $\Gamma$. Additional details of the setup are described elsewhere \cite{Ebert2009,Keim2007}. \\
We follow the well established data evaluation scheme described in \cite{Keim2004}. The harmonic part of the systems potential energy is bilinear in the displacements $\vec u$ times the ``spring constant'' given by \cite{Ashcroft1976}
\begin{equation}
	U = \frac{1}{2}\sum_{\vec q, \alpha, \beta} u_\alpha^*(\vec q) D_{\alpha \beta}(\vec q) u_\beta(\vec q) \,. \label{eq:harmonic_energy}
\end{equation}
For the monolayer the dynamical matrix $D_{\alpha, \beta}(\vec q)$ is a $2 \times 2$ matrix and describes the elastic coupling between particles. It is given by a sum of the second derivatives of the interparticle potential and $\alpha,\beta \in 1,2$ index the two spatial dimensions. Elasticity naturally originates from the spring constants of the system. These are the eigenvalues $\delta_s$ of the dynamical matrix we are interested in. \\
The displacement $\vec u_i(t) = \langle \vec r_i \rangle - \vec r_i(t)$ of particle $i$ is taken relative to its equilibrium position and its Fourier component is given by $\vec{u}(\vec{q}) = \sum_{j=1}^{N} e^{i\vec{q}\cdot\langle\vec{r}_j\rangle} \vec{u}_j(t)$. We calculate those positions to be the center of mass of the trajectory $\langle\vec r_i\rangle = 1/M \cdot \sum_{t}^{M} \vec r_i(t)$, where the average is taken over a time interval $\Delta t$ significantly longer than any short time relaxation process. In the equation, $M$ corresponds to the number of snapshots during the time interval $\Delta t$ and in an amorphous solid those time averaged positions converge to the quasi-equilibrium positions. The expression for $\langle\vec r_i\rangle$ holds as long as $\Delta t$ is smaller than the relaxation time $\tau_\alpha$ and particles are confined in their ``cage''. $\tau_\alpha$ was extracted from MSD measurements for high $\Gamma$. \\
In an amorphous system we cannot calculate the dynamical matrix $D_{\alpha, \beta} (\vec q)$ a priori, but using the equipartition theorem we can measure it's eigenvalues $\delta_s$. The equipartition theorem reads $\langle U \rangle = N \cdot k_B T$ where $N$ is the number of particles, each with two translational degrees of freedom. With Eqn. [\ref{eq:harmonic_energy}] it follows
\begin{equation} \label{eq:equipartition}
  k_BT = \delta_s(\vec q) \cdot \left\langle \left|u_s(\vec q)\right|^2\right\rangle .
\end{equation}
The subscript $s$ denotes polarizations $\parallel,\perp$ and depends on the relative orientation of $\vec u$ and $\vec q$. For $\vec q \parallel \vec u$ and $\vec q \perp \vec u$ we study longitudinal and transverse behaviour, respectively. $\delta_s(\vec q)$ is termed ``dispersion relation'' but we keep in mind that in our system, oscillatory motion is overdamped due to the surrounding solvent. By applying this method, we implicitly assume the investigated modes to be plane waves. This holds in the long wavelength limit corresponding to classical continuum elasticity theory \cite{Landau1991}. There, the 2D system is considered to be a homogeneous structureless solid. Theoretical background on this approach can be found in \cite{Klix2012}. \\

From the dispersion relations, the elastic constants can be derived in the long wavelength limit \cite{Gruenberg2004}. In a two dimensional isotropic solid, the elasticity tensor $C_{\mu\nu\sigma\tau}$ possesses only two independent elements. These two elements can be expressed via the Lam\'{e} coefficients $\mu$ and $\lambda$ of continuum elasticity theory. We extract the Lam\'{e} coefficients from the ``dispersion relation'' as:
\begin{eqnarray}
  \frac{k_BT}{a_0^2\,(2\mu + \lambda)}  & = & \lim_{\vec q \rightarrow 0} [\,\frac{1}{q^2} \langle | u_\parallel(\vec q) |^2 \rangle\,] \, ,\label{eq:elastic_constants1} \\
  \frac{k_BT}{a_0^2\,\mu} & = & \lim_{\vec q \rightarrow 0} [\,\frac{1}{q^2} \langle | u_\perp(\vec q) |^2 \rangle\,] \, .\label{eq:elastic_constants2}
\end{eqnarray}
Here, $a_0$ is the mean interparticle distance, defining the area density as $n = 1/a_0^2$. In two dimensions, $\mu$ gives the shear modulus and $\mu + \lambda = B$ the bulk modulus. As we extract the elastic constants for $\vec q \rightarrow 0$, we are only accounting for long wavelength modes in elastic continuum theory, which justifies our approach even in disordered materials. \\

\begin{figure}
\centerline{\includegraphics[width=.45\textwidth]{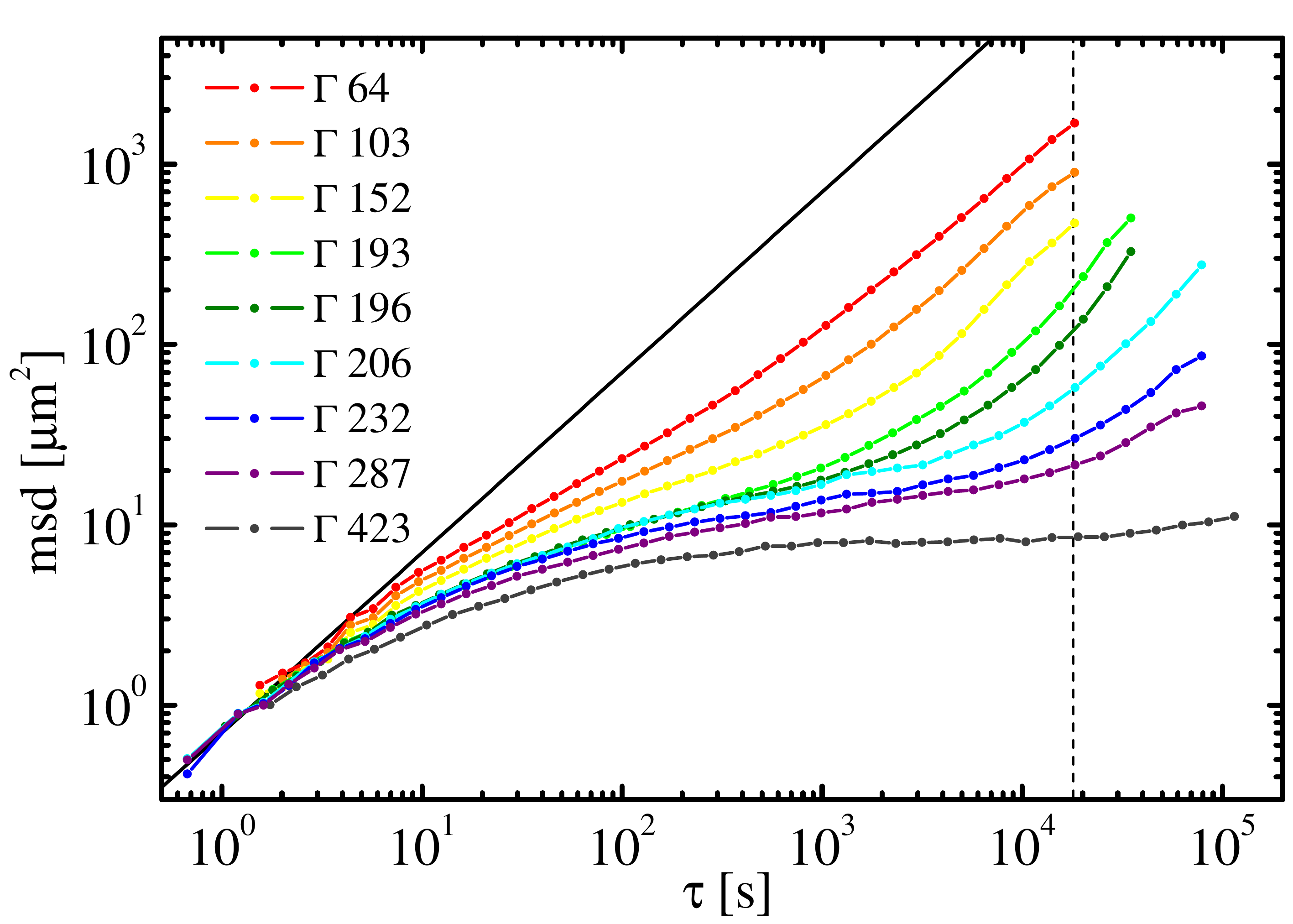}}
\caption{\label{figure_msd} Mean square displacement (msd) for both species of particles at different temperatures. The kink at high temperatures slowly develops into a plateau. However, it is impossible to distinguish unambiguously between solid and fluid state from this dynamical data. The black solid line corresponds to free diffusion and the vertical dashed line to the sampling time $\Delta t$ used in the following analysis to determine the elasticity.}
\end{figure}

Recent work on vibrational properties of colloidal glasses relied on the diagonalisation of the covariance matrix of the displacement field \cite{Ghosh2010,Chen2010,Kaya2010}. From the spectrum of eigenvalues, this two dimensional principal component approach derives the density of vibrational states. In all cases (hard spheres and soft PNIPAM particles) an anomalous low frequency behavior was found, which was connected to the boson peak known from scattering experiments in molecular or polymeric glasses \cite{Malinovsky1986,Sokolov1995}. This excess of modes was attributed to the non-isotropy of disordered systems at short to intermediate lengthscales. In atomic and molecular glasses, recent progress showed that long wavelength modes undergo a crossover from propagation to localization due to diffraction upon decreasing the wavelength \cite{Ruffle2006,Courtens2003,Ruffle2003,Allen1999,Hafner1983}. By analyzing the Eigenvector structure of the covariance matrix, the spatial features of the corresponding modes could be visualized in the colloidal case \cite{Ghosh2010,Chen2010,Kaya2010}. It was concluded that the low frequency spectrum consisted mainly of localized ``swirling'' type of modes. Since this does not go along with our approach of plane waves in a continuous medium, we study the structure of modes via principal component analysis as well at the end of the manuscript. From the particle displacements $\vec u(t)$, we compute the covariance matrix
\begin{equation}\label{eq:covariance_matrix}
	C_{ij} = \langle u_i(t) u_j(t) \rangle_t \, ,
\end{equation}
where $i$ and $j$ correspond to all components of all particle displacements. By diagonalisation we obtain the spectrum of eigenvalues $c^\star$. Following \cite{Ghosh2010,Kaya2010,Chen2010} we use the inverse as frequencies of the modes in our systems, providing a mapping $\omega^2 =  1/{c^\star}$. The corresponding Eigenvectors reveal the structure of those modes we are interested in. In the calculations, we follow the procedures described in \cite{Chen2013}. Because principal component analysis requires large statistics, we use $1.24\cdot 10^5$ frames for constructing the covariance matrix. At this number the results have statistically converged.

\section{Results}

To start off, we present the measured mean squared displacement (msd) $\langle \Delta r^2(\tau)\rangle = \left\langle \frac 1 N \sum_i^N (\vec r_i(t+\tau)-\vec r(t))^2 \right\rangle$ of the particles in Fig. \ref{figure_msd} for different system temperatures. Here, $N$ presents the total number of particles which are located at positions $\vec r_i$ at time $t$. It clearly shows the expected slow down of dynamics. At high temperatures, the data show a mere kink which slowly develops into a plateau for intermediate lag times $\tau$. For longer times $\tau_\alpha$, the dynamics exhibits (sub) diffusive behavior corresponding to alpha relaxation processes. This becomes clear by comparing to the solid black line with slope one, marking free diffusion. It is noteworthy that from this dynamic data alone, the exact temperature of a transition from the fluid state to a solid (amorphous) state can hardly be identified.\\

\begin{figure}[t]
\centerline{\includegraphics[width=.44\textwidth]{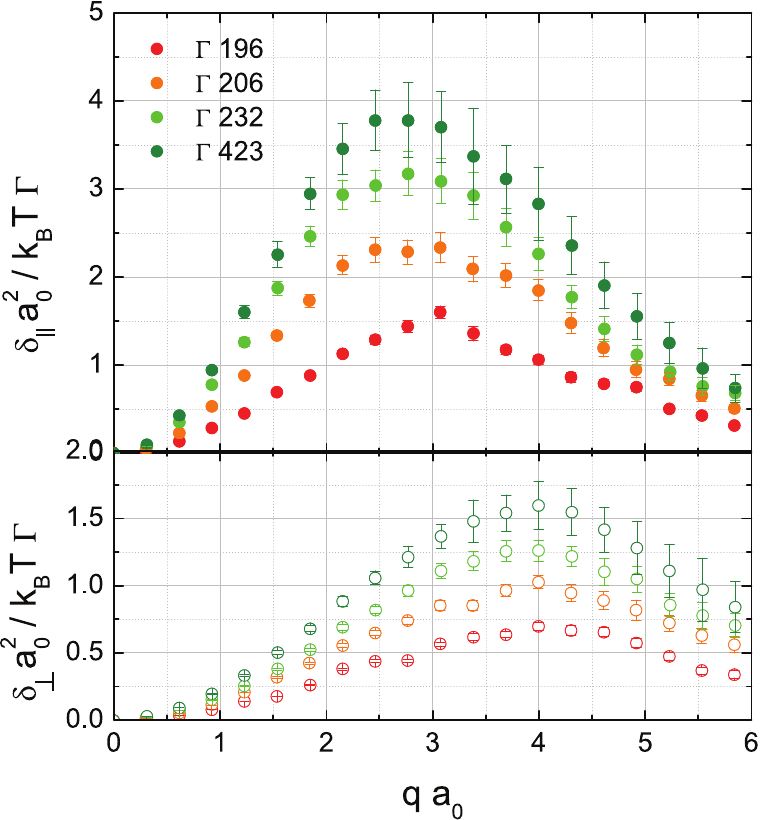}}
\caption{\label{fig:disprel}Dispersion relations of longitudinal and transverse excitations rescaled with the linear dependence on $\Gamma$ expected in harmonic systems (filled and empty symbols in top and bottom plot, respectively). The curves for different temperatures do not collapse, which can only be explained if structural rearrangement occurs. Due to the better optical resolution only trajectories of big particles are analyzed here and in the following plots.}
\end{figure}

Therefore we focus on spectroscopy of acoustic excitations. Fig. \ref{fig:disprel} shows the measured normalized ``dispersion relations'' for different interaction parameters analyzed from trajectories of big particles. Filled and empty symbols represent spring constants for longitudinal and transverse waves, respectively. The curves are averaged over different directions of $\vec q$, taking advantage of the the isotropy of the system. The error bars contain contributions from the finite optical resolution of the microscopy technique (assumed to be $\approx 100~\textrm{~nm}$), which affects the amplitudes of modes in Fourier space. As expected, they show an initial quadratic growth (in agreement with Debye behavior) and exhibit a maximum near $q = \pi$ where the system is stiffest and propagation of waves fades out.\\

\begin{figure}[t]
\centerline{\includegraphics[width=.45\textwidth]{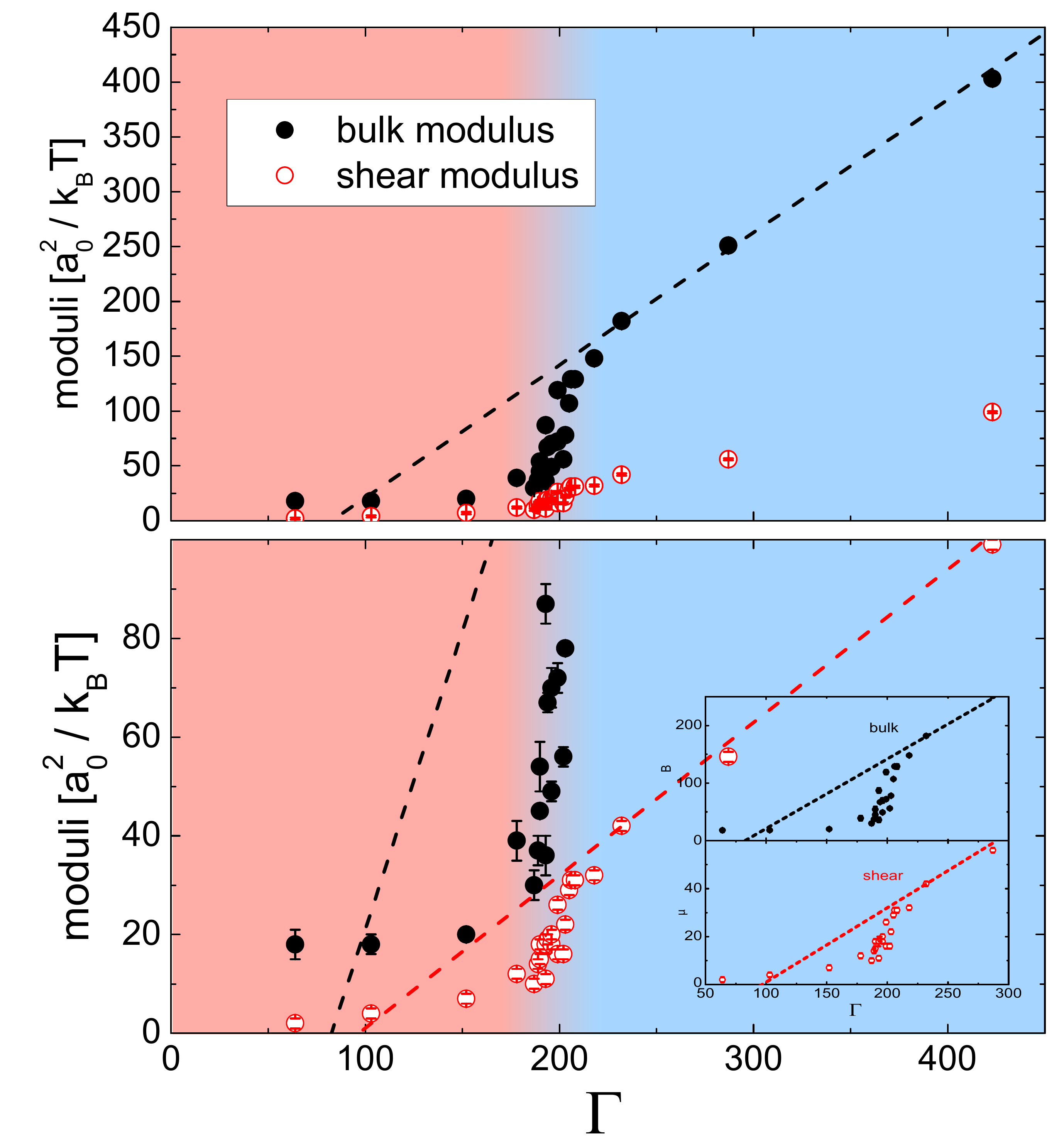}}
\caption{\label{fig:moduli}Temperature dependent bulk- and shear modulus. The data suggest a jump of moduli at the transition temperature, which is located at $\Gamma_T = 195\pm 5$. The lower plot zooms in on the ordinate. In the solid, the moduli grow linearly due to increased pressure indicated by dashed lines and we extrapolate this linear behaviour down to the fluid. The inset shows a zoom of ordinate and abscissa: the data are not compatible with a continuous increase of elasticity as e.g. given by the dotted lines.}
\end{figure}

Usually, the growing amplitude of the curves with increasing $\Gamma$ reflects the stiffening of the system upon cooling.
In a harmonic crystal, the ``dispersion relation'' scales directly with the coupling parameter since the spring constants grow quadratically with $H^2$ and thus go linear with $\Gamma$ \cite{Keim2004}.
Accordingly, the curves for different temperatures should collapse onto one single master curve if divided by the interaction strength. Interestingly, we do not find this scaling in the amorphous solid. Besides the magnetic field, $\Gamma$ only depends on intrinsically fixed parameters as laboratory temperature, magnetic susceptibility or area density. Since the particle-particle interactions (and therefore the ``spring constants'') depend crucially on the respective distance, a subtle change in structure might explain this behavior. With increasing interaction parameter, the particles try to maximize their distances at constant Voronoi area in order to minimize energy. \\
Because in the disordered state the equilibrium particle positions are not fixed for all times, the structure of the system might adapt to the increased pressure without affecting the macroscopic area density. This effect is barely visible in the structure factor. \\
On a side note, the growth of the dispersion amplitude mirroring the increasing rigidity tackles another unique feature of our model system. Since the pair potential is purely repulsive and the system is investigated at constant volume (and particle number), the ensemble is intrinsically under pressure within the boundaries given by the sample cell. Since pressure equals an energy density (that is, energy per area in 2D), the pressure is given by the control parameter $\Gamma$, too, such that temperature and pressure can not be varied independently. A quantitative analysis (see Fig.~\ref{fig:moduli}) shows that the value of the dimensionless pressure given by $\Gamma$ is of the same order of magnitude compared to dimensionless elastic moduli. Such pressure affects the sound velocities given by the slope of the dispersion relation. \\

\begin{figure}[t]
\centerline{\includegraphics[width=.45\textwidth]{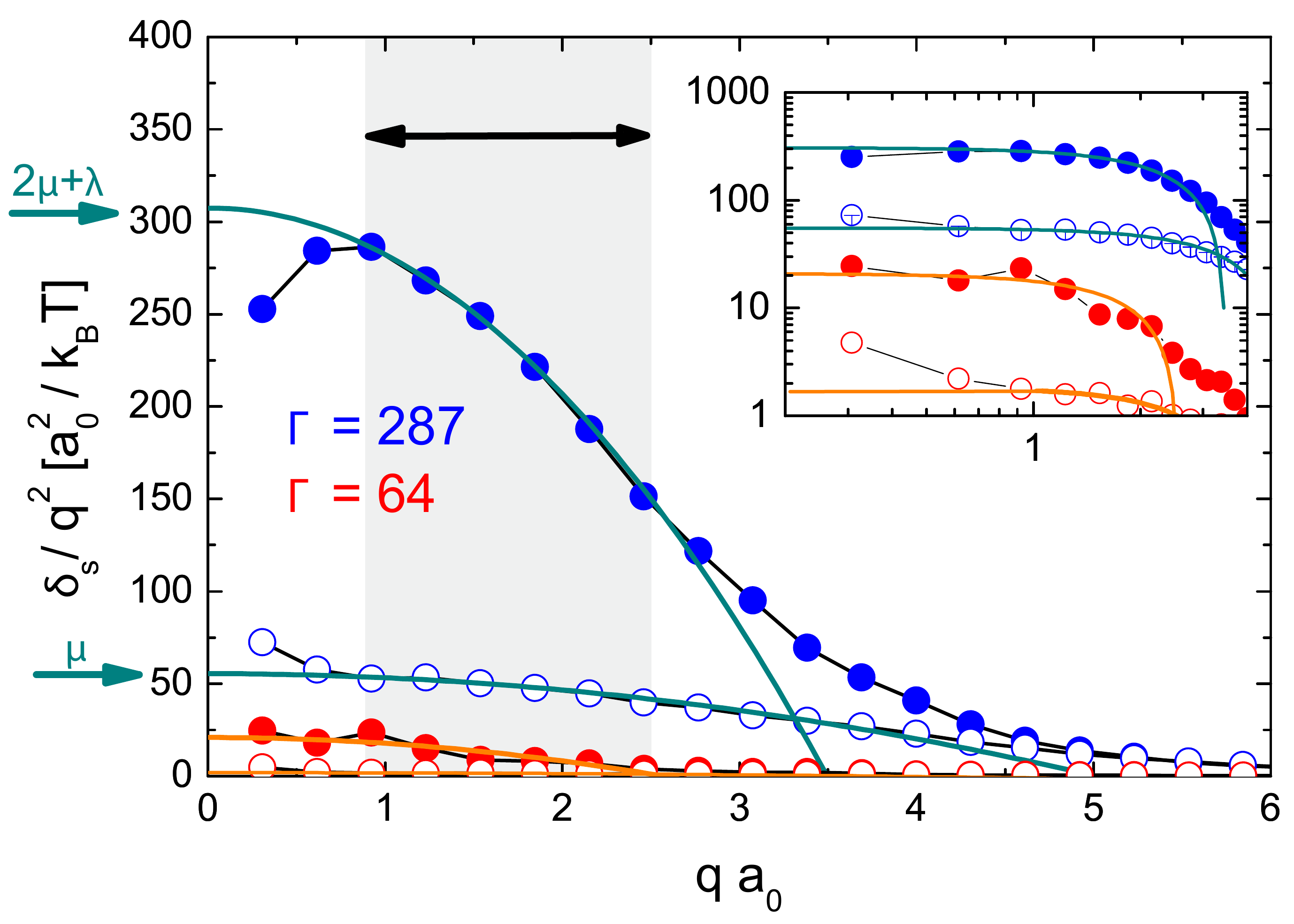}}
\caption{\label{fig:extrapolation} $\mu$ and $2\mu + \lambda$ are given by the intercept of an extrapolation (indicated by green arrows) derived from the longitudinal and transverse dispersion bands (filled and empty black circles) for a solid (blue) and fluid (red) sample. The gray shaded region is used for fitting the data and the inset shows the same plot with logarithmic scaling.}
\end{figure}
Now, we turn to the main results, namely the elastic constants derived from the ``dispersion relation''.
Fig.~\ref{fig:extrapolation} shows exemplary the dispersion relation for $\Gamma = 64$ and $\Gamma = 287$ divided by $q^2$ according Eqn.~\ref{eq:elastic_constants1} and \ref{eq:elastic_constants2}. The solid curves are fits to the data to get the elasticity indicated by green arrows. We chose an intermediate regime ($0.8 < qa < 2.5$) for the extrapolation $\vec q \rightarrow 0$ in the shaded region where the data fits best. While in \cite{Gruenberg2004} a linear fit was used to get the intercept we here expand $\frac{|\sin(q)|^2}{q^2} = 1 - \frac{q^2}{3} + O(q^4)$ and fit quadratically. 30 measurement runs were performed at various temperatures. Due to prior work \cite{Klix2012}, the transition was expected at $180 \lesssim \Gamma \lesssim 240$. Therefore, we focused on good temperature resolution in that temperature range, resulting in high quality data each with measurement length of at least $8\cdot 10^4$ seconds. The result is shown in Fig. \ref{fig:moduli}. The error bars contain contributions from the finite optical resolution and the extrapolation of reduced dispersion relations towards small $q$ as described by Eqn. [\ref{eq:elastic_constants1}] and [\ref{eq:elastic_constants2}]. The statistics for the latter was greatly improved by averaging over different directions of $\vec q$. In the limits of these errors, the shear modulus below $\Gamma \approx 190$ equals zero, regardless of the applied external magnetic field. For increasing temperatures above, the moduli grow linearly, reflecting the linear ``stiffening'' of the interaction potential. With the amount of data, a zoom in on the shear modulus (shown in the lower plot) clearly reveals an abrupt rise of $\mu$, consolidating the results of simulation \cite{Klix2012} and delivering the first experimental evidence for a discontinuous behavior of the shear modulus at the glass transition. This jump separates the fluid phase from the amorphous state, as indicated by the shading from red to blue.
Averaging about the data in the transition region, we find $\Gamma_T = 195\pm 5$. The scattering of the data is attributed to dynamical heterogeneity present in our system \cite{Mazoyer2011} which have been shown to be correlated with elastic heterogeneity \cite{Wagner2011}.
The dashed lines are fits to the moduli in the region $220 < \Gamma < 450$. For the amorphous system we find a slope of $\mu \simeq 0.31 \cdot \Gamma$ and $ B = \lambda + \mu \simeq 1.21 \cdot\Gamma$. This compares to measurements for dipolar monodisperse \emph{crystals}, where we found $\mu = 0.35 \cdot\Gamma$ and $B = 3.46 \cdot \Gamma $ with a melting temperature at $\Gamma = 60$, see Ref. \cite{Gruenberg2004}. The latter is consistent with a thermodynamic calculation for a dipolar crystal at $T = 0$ which yields a ratio of bulk to shear modulus of $B/\mu = 10$ \cite{Kusner1993}. An increased ratio $B/\mu$ compared to crystals is proposed by nonaffine displacements affecting mainly the shear modulus of systems with short range interaction \cite{Zaccone2014}. For the dipolar monolayer the ratio $B/\mu \simeq 3.5$ is decreased, dominantly due to a reduced bulk modulus. \\

Concerning the height of the jumps we are not aware of predictions of universal behavior of elasticity in thermal amorphous systems as e.g. the famous $16\pi\approx 50.26$ for Youngs modulus in 2D melting \cite{Chaikin2000}, but the measured values in glass and crystal are of the same order. For the jump in the bulk modulus $B_{solid} = m \cdot B_{fluid}$ a factor $m = 1.7$ is calculated by mode coupling theory for hard spheres in 3D \cite{Fuchs1992} and implicitly given in \cite{Bayer2007} where it is shown that 3D and 2D systems behave comparably. For the dipolar system at hand we observe a factor $m \approx 6$ for the glass and $m \approx 4$ for a crystal \cite{factorjump}. In comparison to hard discs we attribute this increased factor to the softness of the dipolar potential. This discontinuous behaviour distinguishes the glass transition from jamming phenomena, for which the modulus was found to grow continuously from zero. This is encompassing granular systems \cite{OHern2003,Wyart2005}, percolating gels with attractive interactions \cite{Winter1986,Prasad2003} and furthermore systems exhibiting friction between particles \cite{Menut2012,Still2014}. This might be an ubiquitous difference between thermal and granular systems. \\

While a recent discussion about 2D and 3D glass claims to find fundamental differences \cite{Flenner2015b} we expect the qualitative behavior to be robust against changes in dimensionality. In 2D crystals the existence of Peierls instabilities \cite{Peierls1934} or Mermin-Wagner fluctuations \cite{Mermin1966,Mermin1968} as long wavelength density fluctuations (of the order of system size) is well established. Peierls argument using relative distance fluctuations is not based on periodicity (just a typical particle distance is needed unlike e.g. in a gas) and thus we expect such long wavelength density fluctuations (with diverging amplitude in the thermodynamic limit) also in 2D glass. Using local coordinates namely cage-relative displacements (called modified dynamic lindemann-parameter for 2D crystals \cite{Zheng1998,Zahn2000}) as suggested in \cite{Mazoyer2009,Mazoyer2011} would solve most of the discrepancies found in \cite{Flenner2015b}. In the present analysis global coordinates are used and Mermin-Wagner fluctuations are expected to suppress the elastic response in the limit of infinite time scales. Thus the time interval $\Delta t$ to determine the quasi-equilibrium position has to be chosen carefully.

\begin{figure}[t]
\centerline{\includegraphics[width=.45\textwidth]{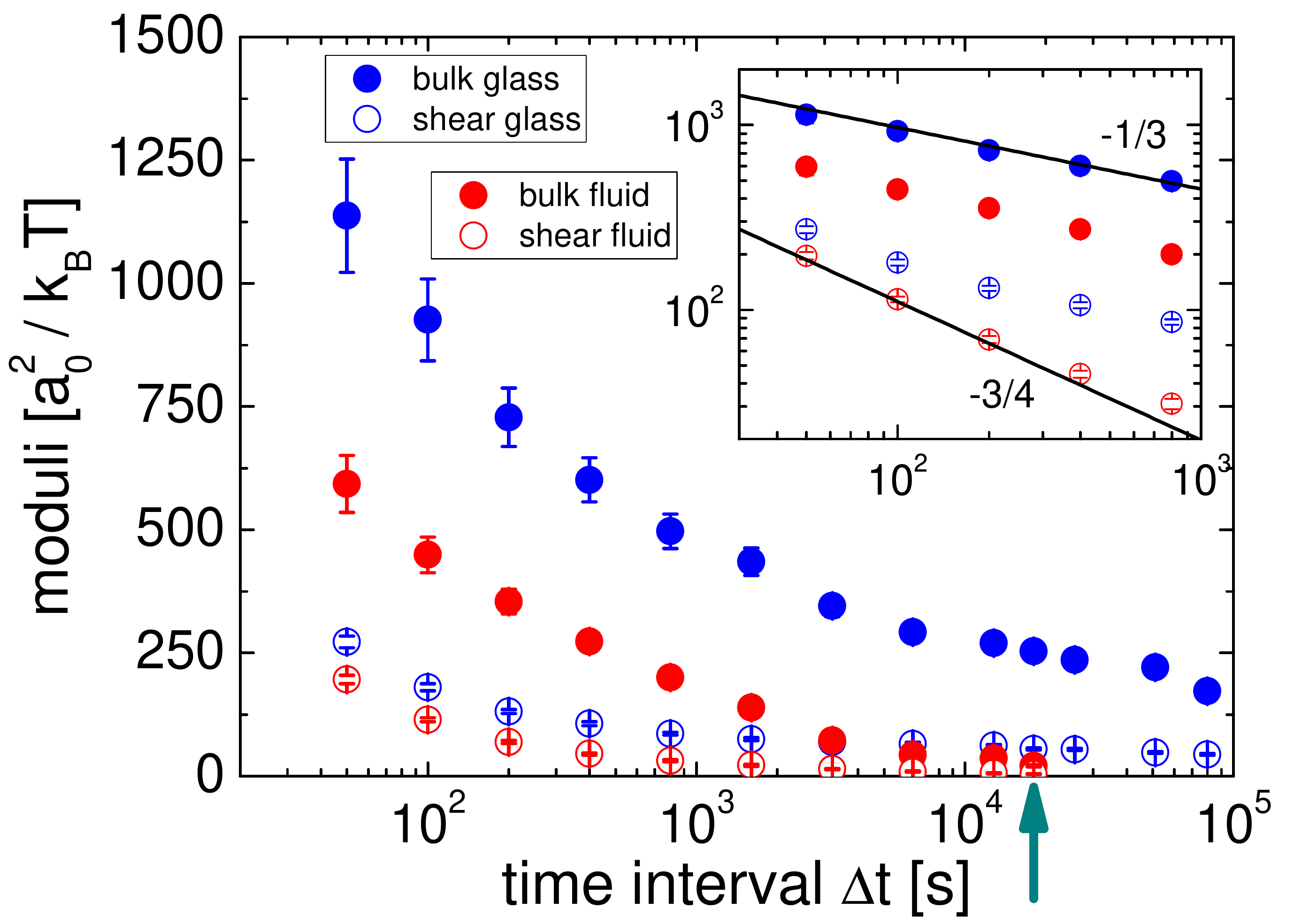}}
\caption{\label{fig:FT}Finite time effects investigated by varying the time interval $\Delta t$ used to determine the particle ``equilibrium position''. On the glassy side, the shear modulus does not decay to zero even for the longest integration times. Data of the system at $\Gamma = 287$ in the solid (blue) and $\Gamma = 103$ in the fluid (red) while the bulk and shear moduli are marked by filled and empty symbols. The inset shows the high frequency region in the frozen state, which suggests a power law behavior. Upper and lower black lines are functions with exponents $-\frac 13$ and $-\frac 34$.}
\end{figure}

\begin{figure*}
\begin{center}
\centerline{\includegraphics[width=.8\textwidth]{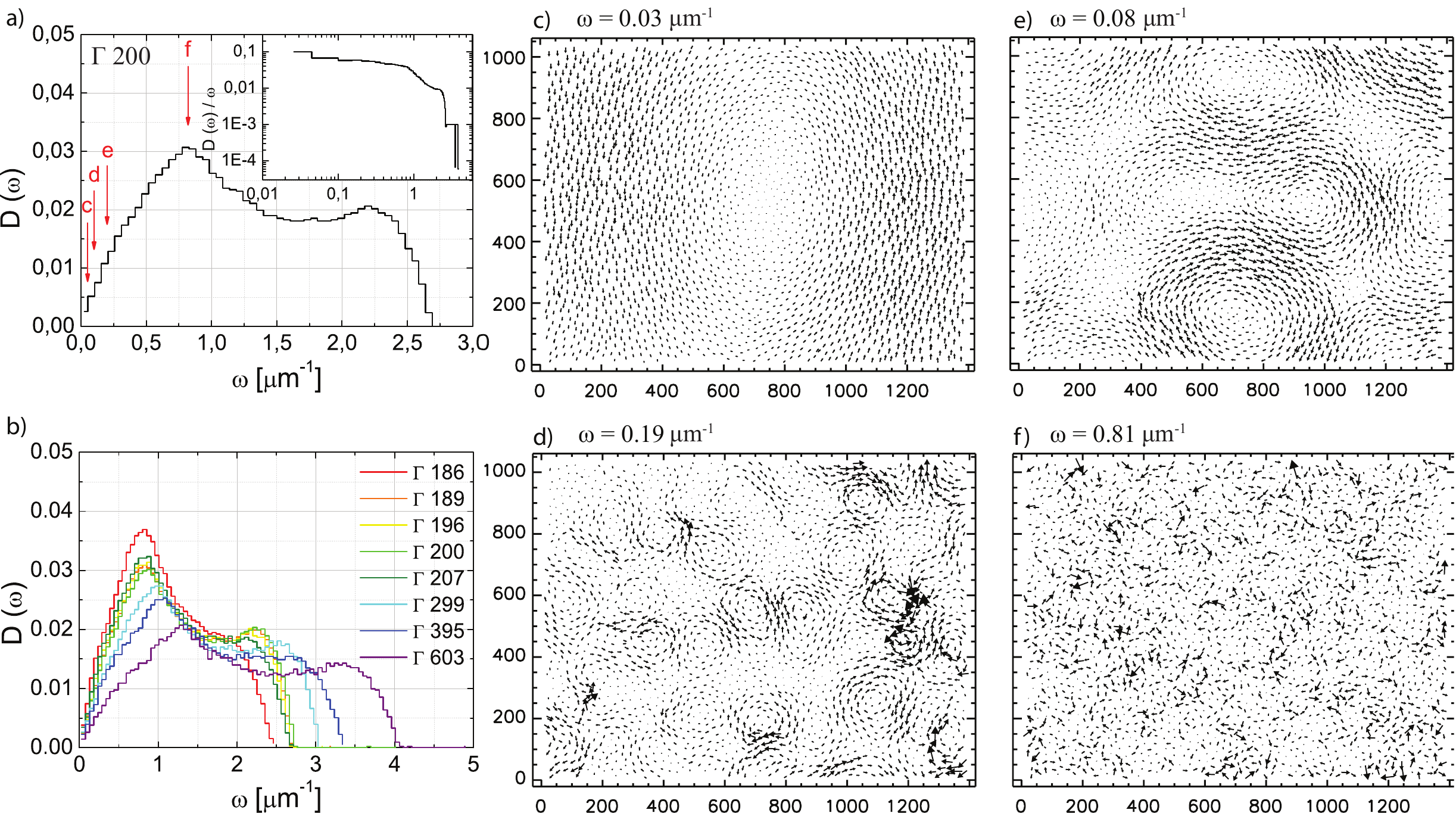}}
\caption{\label{fig:modestructure} Graph a shows the spectrum of eigenvalues of the displacement field covariance matrix of big particles. By scaling with $\omega$, the inset reveals convincing Debye behavior. The red arrows indexed c-f indicate positions in the spectrum where the mode structure is shown in graphs  c-f. The corresponding ``frequencies'' are given above the respective graph. For details, see text.}
\end{center}
\end{figure*}
In Fig. \ref{fig:FT}, the finite time effects of the evaluation procedure are shown. By changing the averaging time $\Delta t$ determining the quasi-equilibrium positions, different frequency regimes are probed. Short averaging times yield a significant increase in elastic moduli, reflecting the expected stiffening at high frequencies even for fluids \cite{Jeong1987}. For low frequencies (large $\Delta t$), in contrast, the shear modulus of the fluid state decays to zero, while in the glassy state, it decays into a plateau, even for times up to $10^5$ seconds. This stability of the shear modulus is remarkable, because other quantities measuring the system dynamics already show signs of relaxation. This averaging time used in Fig.~\ref{fig:moduli} is $\Delta t = 1.8\cdot 10^4$ indicated by the green arrow in Fig.~\ref{fig:FT} (limited by the longest accessible time from the fluid data). Comparing with Fig.~\ref{figure_msd} this is well after the onset of the alpha relaxation time $\tau_\alpha$ for the datasets in the vicinity of the transition. In contrast to the shear modulus, the bulk modulus is much more sensitive to increased dynamics, as it does not show a well developed plateau for low frequencies. This is attributed to the fact that transverse shear waves have much larger amplitudes than longitudinal compression waves, such that distortions of the wavefront affect properties derived from the latter much more strongly. This striking behavior proves the shear modulus to be a good quantity to investigate the glass transition. The inset finally shows a log-log plot of the high frequency regime for the frozen state. For about two decades, both moduli show a power law behavior. The blue lines are functions with exponents of $-\frac{1}{3}$, describing high frequency bulk and shear properties, respectively. This power law behavior challenges theoretical descriptions.\\

In the last section we turn to the analysis of the shape of the excitations. Fig. \ref{fig:modestructure}\,a) shows the spectrum of eigenvalues $D(\omega)$ of the displacement field covariance matrix $C_{ij}$ for the system at $\Gamma = 200$, a temperature slightly below the glass transition temperature. The most \textit{principal} components (carrying the most weight) correspond to low ``frequencies'' due to the mapping $\omega \propto 1/\sqrt{c^\star}$. Because $C_{ij}$ has the dimensions of square meters as it measures amplitudes, the ``frequencies'' have the unit $\mu m^{-1}$. Only a valid mapping of amplitudes, wavelength, and frequencies (implicitly using a dispersion relation) connects a length scale given by the eigenvalues of the covariance matrix to frequencies used in the density of states $D(\omega)$. An interesting feature of $D(\omega)$ is the double peak structure, centering at $\omega \approx 0.8~\mu m^{-1}$ and $\omega \approx 2.1~\mu m^{-1}$. We interpret this as a reminiscence of van-Hoove type singularities as they appear e.g. in crystals due to the vanishing sound velocity at the Brillouin zone edge \cite{Chen2013}. Naively one could argue that first and second peak represent longitudinal and transverse modes respectively, but since PCA is a priori insensitive to polarizations the connection is less trivial. Nevertheless, Fig.~\ref{fig:modestructure}\,b) shows the density of states as function of temperature. The peaks shift to lower frequencies with increasing temperature (decreasing interaction strength) reflecting a softening of the system as expected. Note that the first peak grows at the expense of the second upon approaching the fluid phase. This resemblance with a density of states known from a crystalline solid also shows up in the reduced density of states $D(\omega)/\omega$ which is shown in the inset of Fig. \ref{fig:modestructure}\,a). Scaling with $\omega$ yields a constant value for the low frequency region up to the first peak. This is in accordance with the Debye model $D(\omega) \propto \omega^{d-1}$, where $d$ is the dimensionality of the investigated system. In the Debye model, it is plane wave excitations which make up the low frequency part of the spectrum. Putting issues regarding the amplitudes frequency mapping aside, the mode structure is clearly resolved by the eigenvectors of the most principal components of the displacement field. For decreasing wavelength, these are shown in graphs \ref{fig:modestructure}\,c-f with the corresponding frequency noted above. Those frequencies are also indicated by red arrows in Fig. \ref{fig:modestructure}\,a). While \ref{fig:modestructure}\,e,d) show some kind of Lissajous-pattern, for modes of a given frequency but different phases, they start to show incoherence for higher frequencies indicated by a random distribution of arrows. The coherence length of the imaged virtual motion gets notably smaller, reflecting the growing influence of disorder. This culminates at frequencies accounting to the first peak, where the mode structure appears almost disordered in Fig. \ref{fig:modestructure}\,f). Nonetheless, care has to be taken since PCA was shown to generate mode pattern from mixed states at higher frequencies; the low frequency modes on the other hand are expected to be pure states \cite{Maggs2015}. In Fig. \ref{fig:modestructure}\,c) one can clearly verify the plane wave character of the excited states in the low frequency limit. Accordingly, the ansatz to equally distribute energies via the equipartition theorem on plane wave like modes for $\lim (q \to 0)$ (Eqn. \ref{eq:equipartition}) is reasonable.

\section{Conclusion}

We have experimentally investigated the elastic properties of a colloidal glass former in great detail. By tuning the interactions \textit{in situ}, we have accessed the well defined displacement field in Fourier space at numerous effective temperatures and used the equipartition theorem to calculate the dispersion relations. Scaling with the interaction parameter reveals a hidden structural change upon cooling which is too subtle to be picked up by other functions. In the limit of long wavelength, the dispersion relations yield the elastic moduli of compression and shear, of which particularly the latter is of great interest for an deeper understanding of the glass transition. It has been demonstrated that the shear response unambiguously separates the fluid state from the amorphous solid. The transition itself exhibits a discontinuous behavior, which is in agreement with theoretical predictions and simulations and distinguishes it from granular matter or gelation type scenarios. Finite time analysis reveals a striking stability of the shear-related measurements, which are proven to be decoupled from relaxation phenomena like cage breaking. This makes the shear modulus an exceptionally good quantity to investigate the glass transition.

\begin{acknowledgments}
We are grateful for fruitful discussions with M. Fuchs and W. Schirmacher, P.K. further acknowledges financial support from the Young Scholar Fund (YSF) University of Konstanz.
\end{acknowledgments}

\end{document}